\documentclass{article}

\PassOptionsToPackage{numbers, compress}{natbib}
\usepackage[final]{neurips_2020}
\usepackage[utf8]{inputenc} 
\usepackage[T1]{fontenc}    
\usepackage{hyperref}       
\usepackage{url}            
\usepackage{booktabs}       
\usepackage{amsfonts}       
\usepackage{nicefrac}       
\usepackage{microtype}      
\usepackage{algorithm}
\usepackage{algorithmic}
\usepackage{mathtools}

\usepackage{bm}

\usepackage{graphicx}
\usepackage{float}
\usepackage{xcolor}

\title{FinRL-Meta: A Universe of Near-Real Market Environments for Data-Driven Deep Reinforcement Learning in Quantitative Finance}

\author{    Xiao-Yang Liu$^1$, Jingyang Rui$^2$, Jiechao Gao$^3$, Liuqing Yang$^1$, Hongyang Yang$^1$, \\ \textbf{Zhaoran Wang$^4$, Christina Dan Wang$^5$\thanks{Christina Dan Wang is supported in part by National Natural Science Foundation of China (NNSFC) grant 11901395 and Shanghai Pujiang Program, China 19PJ1408200.},~
, Jian Guo$^6$}\\ 
   $^1$Columbia University; $^2$The University of Hong Kong; $^3$University of Virginia; \\
  $^4$Northwestern University;
  $^5$New York University (Shanghai); $^6$IDEA Research.\\
  \texttt{XL2427@columbia.edu; zhaoranwang@gmail.com; guojian@idea.edu.cn} \\  
       }

\begin{document}

\maketitle

\begin{abstract}

Deep reinforcement learning (DRL) has shown huge potentials in building financial market simulators recently. However, due to the highly complex and dynamic nature of real-world markets, raw historical financial data often involve large noise and may not reflect the future of markets, degrading the fidelity of DRL-based market simulators. Moreover, the accuracy of DRL-based market simulators heavily relies on numerous and diverse DRL agents, which increases demand for a universe of market environments and imposes a challenge on simulation speed. In this paper, we present a FinRL-Meta framework that builds a universe of market environments for data-driven financial reinforcement learning. First, FinRL-Meta separates financial data processing from the design pipeline of DRL-based strategy and provides open-source data engineering tools for financial big data. Second, FinRL-Meta provides hundreds of market environments for various trading tasks. Third, FinRL-Meta enables multiprocessing simulation and training by exploiting thousands of GPU cores. Our codes are available online at \url{https://github.com/AI4Finance-Foundation/FinRL-Meta}. \looseness=-1
  
\end{abstract}

\section{Introduction}

In quantitative finance, market simulators play important roles in studying the complex market phenomena and investigating financial regulations \cite{raberto2001agent, mizuta2016brief}. Compared to traditional simulation models, deep reinforcement learning (DRL) has shown huge potentials in building financial market simulators through multi-agent systems \cite{lussange2021modelling}. However, due to the high complexity of real-world markets, raw historical financial data involve significant noise and may not reflect the future of markets. This issue usually degrades the fidelity of DRL-based simulation. Moreover, in reality, there are innumerable participators that impact together on the market. To better simulate the market, numerous and diverse DRL agents are needed to represent those participators with different aims and strategies.

Recently, researchers have explored various applications of DRL in quantitative finance \cite{lussange2021modelling, liu2021finrl, karpe2020multi, pricope2021deep}. Lussange et al. \cite{lussange2021modelling} have proposed a market simulation model using multi-agent reinforcement learning. Although it has shown the feasibility of DRL-based market simulation, only a few DRL agents are used. The potentials of DRL-based market simulators are not fully explored yet. The FinRL framework \cite{liu2021finrl} has proposed a DRL framework as a full pipeline of developing trading strategies and is growing an open-source community \textit{AIFinannce} that can contribute diverse DRL agents. However, it \cite{liu2021finrl} focuses on developing trading strategies instead of building market simulations.

In this paper, we develop a FinRL-Meta framework that is a universe of near real-market environments for data-driven financial reinforcement learning. First, we apply the DataOps paradigm \cite{DataOps} to the data engineering pipeline, providing agility to agent deployment. We offer a unified and automated data processor for data accessing, data cleaning and feature engineering. Second, we build hundreds of near real-market DRL environments for various trading tasks such as high-frequency trading, cryptocurrencies trading, stock portfolio allocation, etc.. The environments are directly connected to our data processor. High-quality large datasets can be generated efficiently and encapsulated into our environments. Third, to accelerate the training process of DRL agents in large datasets, we utilize thousands of GPU cores to perform multiprocessing training. \looseness=-1



\begin{figure}
\centering
\includegraphics[scale = 0.2]{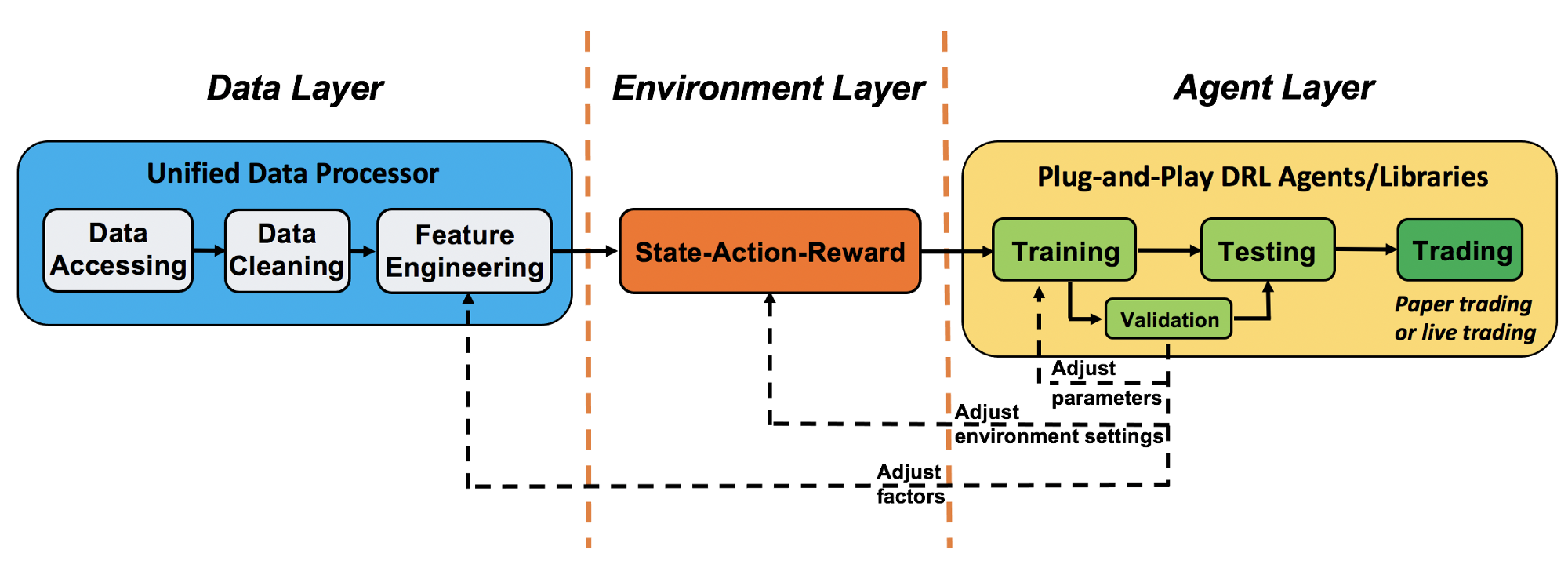}\vspace{-0.1in}
\caption{Overview of FinRL-Meta.}
\vspace{-0.1in}
\label{fig:finrl-meta overview}
\vspace{-0.1in}
\end{figure}

\vspace{-0.15in}
\section{Proposed FinRL-Meta Framework}

\begin{table}[t]
\resizebox{\textwidth}{13mm}{
\begin{tabular}{c c c c c c}
\hline
\textbf{Data Source} & \textbf{Type} & \textbf{Range and Frequency} & \textbf{Request Limits} & \textbf{Raw Data} & \textbf{Preprocessed Data}\\
\hline
Yahoo! Finance& US Securities& Frequency-specific, 1min& 2,000/hour& OHLCV& Prices \& Indicators\\
CCXT& Cryptocurrency& API-specific, 1min& API-specific& OHLCV& Prices \& Indicators\\
WRDS.TAQ& US Securities& 2003-now, 1ms& 5 requests each time& Intraday Trades& Prices \& Indicators\\
Alpaca& US Stocks, ETFs& 2015-now, 1min& Account-specific& OHLCV& Prices \& Indicators\\
RiceQuant& CN Securities& 2005-now, 1ms& Account-specific& OHLCV& Prices \& Indicators\\
JoinQuant& CN Securities& 2005-now, 1min& 3 requests each time& OHLCV& Prices \& Indicators\\
QuantConnect& US Securities& 1998-now, 1s& NA& OHLCV& Prices \& Indicators\\
\hline
\end{tabular}}
\caption{Data platforms. OHLCV means open, high, low, close, volume data.}
\vspace{-0.2in}
\label{supported platforms}
\end{table}

\textbf{MDP Model for Trading Tasks}:
We model a trading task as a Markov Decision Process (MDP) $(\mathcal{S}, \mathcal{A}, P, r, \gamma)$ \cite{sutton2018reinforcement}, where $\mathcal{S}$ and $\mathcal{A}$ denote the state space and action space, respectively, $P(s’|s,a)$ denotes the transition probability, $r(s,a)$ is a reward function, and $\gamma \in (0,1) $ is a discount factor. Specifically, the state denotes an observation that a DRL agent receives from a market environment; the action space consists of actions that an agent is allowed to take at a state; the reward function $r(s,a,s')$ is the incentive for agents to learn a better policy. A trading agent aims to learn a policy $\pi(s_t|a_t)$ that maximizes the expected return $R = \sum^{\infty}_{t=0} \gamma ^t r(s_t, a_t)$. 

\textbf{Overview of FinRL-Meta}:
We utilize a layered structure, as shown in Fig. \ref{fig:finrl-meta overview}. FinRL-Meta consists of three layers: data layer, environment layer, and agent layer. Each layer executes its functions and is relatively independent. Meanwhile, layers interact through end-to-end interfaces to implement the complete workflow of algorithm trading. 

\textbf{DataOps for Data-Driven DRL in Finance}:
We follow the DataOps paradigm \cite{DataOps} in the data layer. First, we establish a standard pipeline for financial data engineering, ensuring data of different formats from different sources can be incorporated in a unified RL framework. Second, we automate this pipeline with a data processor, which can access data, clean data and extract features from various data sources with high quality and efficiency. Our data layer provides agility to model deployment. The data sources are shown in Table \ref{supported platforms}.

\textbf{Multiprocessing Training}:
We utilize thousands of GPU cores to perform multiprocessing training, which significantly accelerates the training process. In each CUDA core, a trading agent interacts with a market environment to produce transitions in the form of (state, action, reward, next state). Then all the transitions are stored in a replay buffer and used to update a learner and evaluator. By adopting this technique, we successfully achieve multiprocessing simulation of hundreds of market environments to improve the performance of DRL trading agents on large datasets. 




\textbf{Plug-and-Play}: In the development pipeline, we separate market environments from the data layer and the agent layer. Any DRL agent can be directly plugged into our environments, then trained and tested. Different agents can run on the same benchmark environment for fair comparison.

\textbf{Training-Testing-Trading Pipeline}:
We employ a training-testing-trading pipeline. The DRL agent first learns from the training environment and is then validated in the validation environment for further adjustment. Then the validated agent is tested on historical datasets. Finally, the tested agent will be deployed in paper trading or live trading markets. First, this pipeline solves the information leakage problem because the trading data are generated yet when adjusting agents. Second, a unified pipeline allows fair comparisons among different trading strategies.

\textbf{Supported Trading Tasks}:
We have supported and achieved satisfactory trading performance for trading tasks such as stock trading, cryptocurrency trading, and portfolio allocation. Derivatives such as futures and forex are also supported. Besides, we have supported multi-agent simulation and execution optimizing tasks by reproducing the experiment in \cite{bao2019multiagent}. 

\vspace{-0.1in}
\section{Performance Evaluation}

\begin{figure*}[t]
\centering
\includegraphics[width=0.45\textwidth]{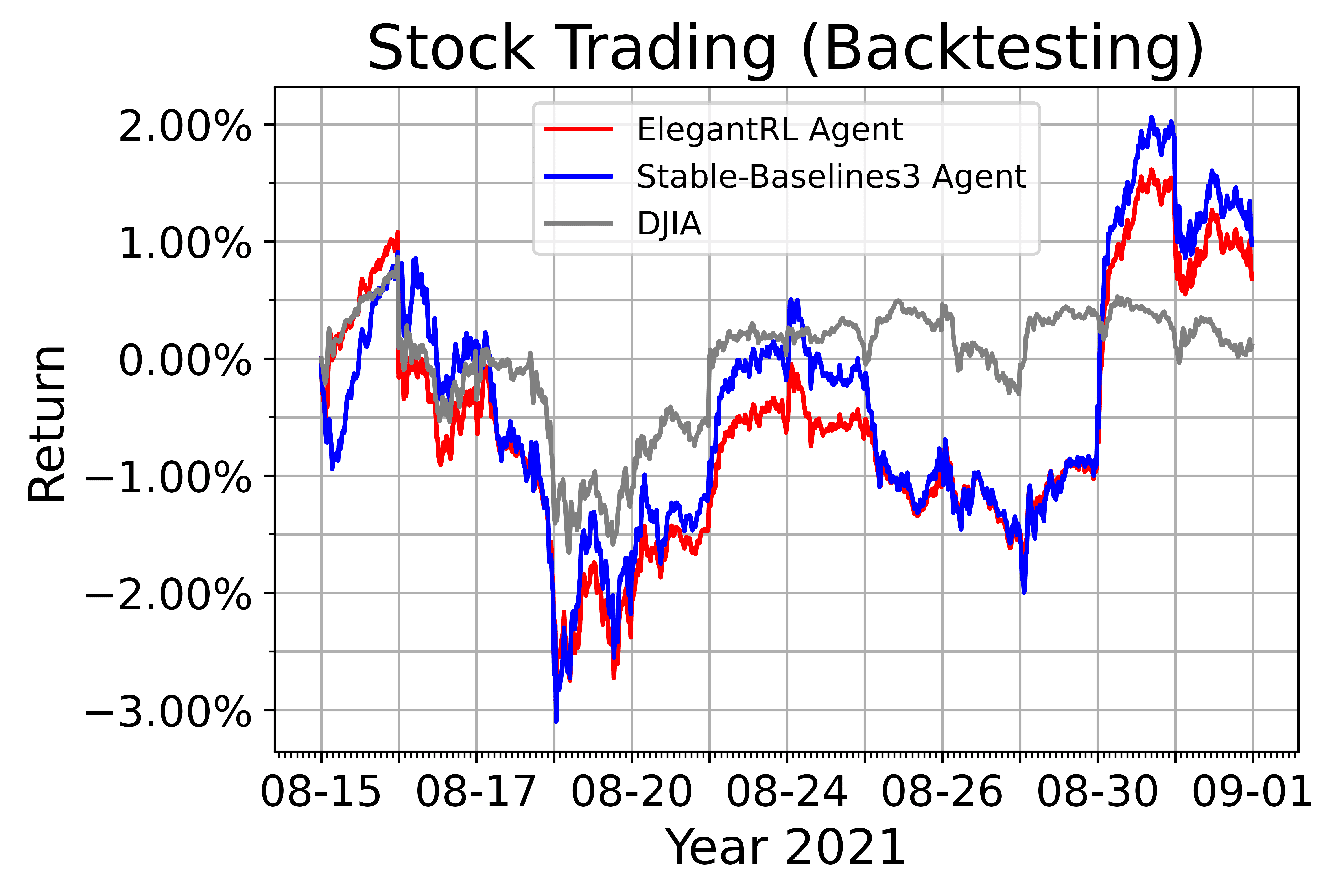}
\includegraphics[width=0.45\textwidth]{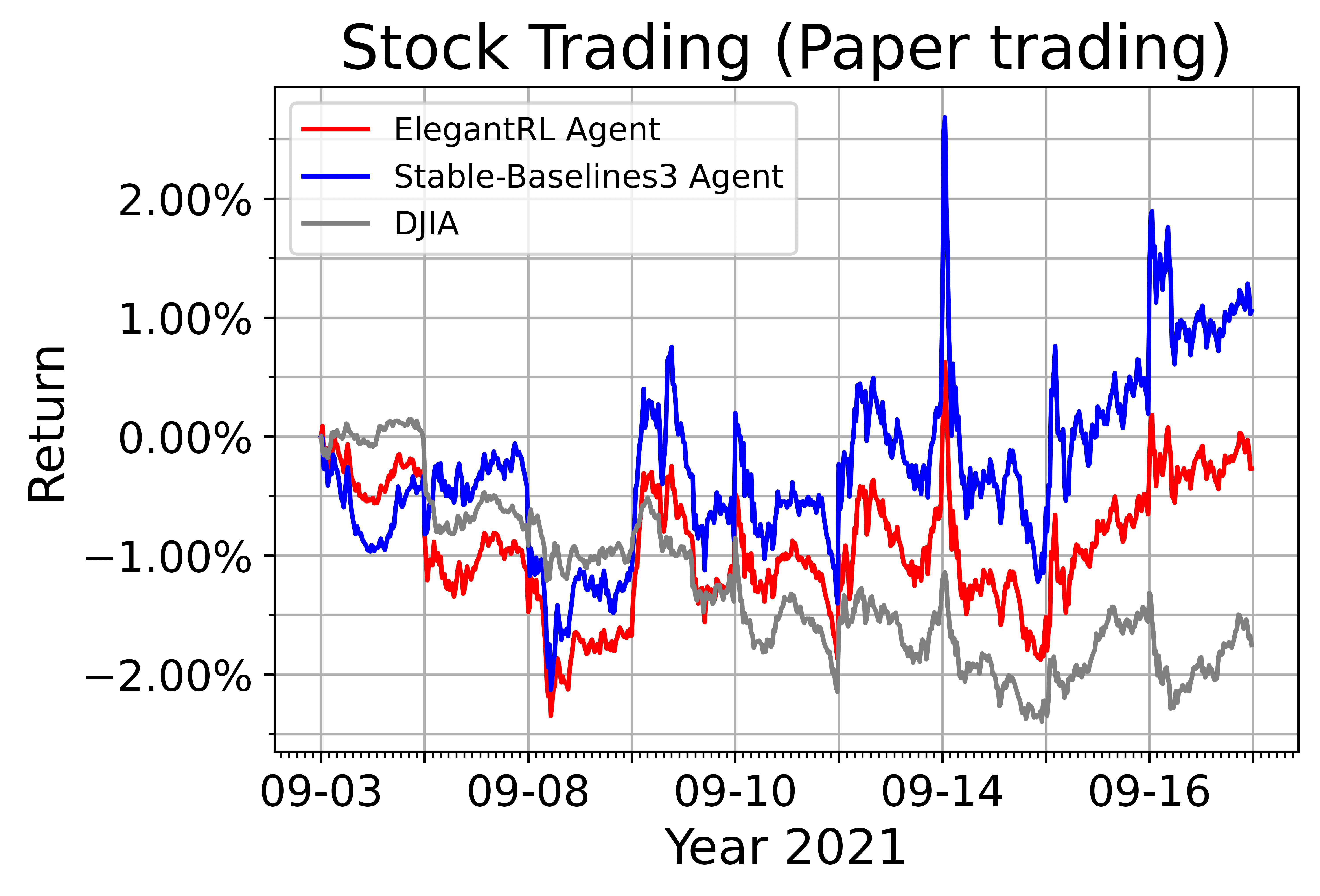}
\caption{Cumulative returns (5-minute) of stock trading in backtesting and paper trading.}
\label{fig:stock_trading}

\end{figure*}

\begin{table*}[t]
\centering
\resizebox{\textwidth}{13mm}{
\begin{tabular} {|l|c|c|c|} 
\hline
&\textbf{ElegantRL} \cite{elegantrl} &\textbf{Stable-baselines3 } \cite{stable-baselines} &\textbf{DJIA}\\
\hline
\textbf{Cumul. return}  &\textcolor{red}{0.968\%} / \textcolor{blue}{-0.652\%} & \textcolor{red}{1.335\%} / \textcolor{blue}{0.191\%} & \textcolor{red}{0.099\%} /
\textcolor{blue}{-1.56\%} \\
\hline
\textbf{Annual return}  &\textcolor{red}{22.425\%} / \textcolor{blue}{-16.746\%} & \textcolor{red}{32.106\%} / \textcolor{blue}{5.492\%} & \textcolor{red}{2.108\%} / \textcolor{blue}{-35.522\%} \\
\hline
\textbf{Annual volatility} & \textcolor{red}{15.951\%} / \textcolor{blue}{14.113\%} & \textcolor{red}{19.871\%} / \textcolor{blue}{15.953\%} & \textcolor{red}{9.196\%} / \textcolor{blue}{9.989\%}\\
\hline
\textbf{Sharpe ratio}  &     \textcolor{red}{1.457} / \textcolor{blue}{-1.399} &   \textcolor{red}{1.621} / \textcolor{blue}{0.447} &    \textcolor{red}{0.289} / \textcolor{blue}{-4.894} \\
\hline
\textbf{Max drawdown} & \textcolor{red}{-2.657\%} / \textcolor{blue}{-1.871\%} &    \textcolor{red}{-2.932\%} / \textcolor{blue}{-1.404\%} & \textcolor{red}{-1.438\%} / \textcolor{blue}{-2.220\%}\\
\hline
\end{tabular}}
\caption{Performance of backtesting  (\textcolor{red}{red}) and paper trading (\textcolor{blue}{blue}) for stock trading.}
\label{table_stock}
\vspace{-0.15in}
\end{table*}

To provide benchmarks for researchers, we will continuously add typical trading tasks with corresponding environments. Here, we show results of stock trading and cryptocurrency trading.

\vspace{-0.15in}
\subsection{Experiment Settings}

\textbf{Stock trading task}: We select the 30 constituent stocks in Dow Jones Industrial Average (DJIA), accessed at the beginning our testing period. We use the Proximal Policy Optimization (PPO) algorithm \cite{PPO_2017} of ElegantRL \cite{elegantrl}, Stable-baselines3 \cite{stable-baselines} and RLlib \cite{liang2018rllib}, respectively, to train agents and use the DJIA index as the baseline. We use 1-minute data from 06/01/2021 to 08/14/2021 for training and data from 08/15/2021 to 08/31/2021 for validation (backtesting). Then we retrain the agent using data from 06/01/2021 to 08/31/2021 and conduct paper trading from 09/03/2021 to 09/16/2021. The historical data and real-time data are accessed from the Alpaca's database and paper trading APIs. 

\textbf{Cryptocurrency trading task}: We select top 10 market cap cryptocurrencies \footnote{The top 10 market cap cryptocurrencies as of Oct 2021 are: Bitcoin (BTC), Ethereum (ETH), Cardano (ADA), Binance Coin (BNB), Ripple (XRP), Solana (SOL), Polkadot (DOT), Dogecoin (DOGE), Avalanche (AVAX), Uniswap (UNI). Tether (USDT) and USD Coin (USDC) are excluded.}. We use the PPO algorithm \cite{PPO_2017} of ElegantRL \cite{elegantrl} to train an agent and use the Bitcoin (BTC) price as the baseline. We use 5-minute data from 06/01/2021 to 08/14/2021 for training and data from 08/15/2021 to 08/31/2021 for validation (backtesting). Then we retrain the agent using data from 06/01/2021 to 08/31/2021 and conduct paper trading from 09/01/2021 to 09/15/2021. The historical data and real-time data are accessed from Binance.

\vspace{-0.15in}

\subsection{Trading Performance}

\textbf{Stock trading}:
In the backtesting stage, both ElegantRL \cite{elegantrl} agent and Stable-baselines3 \cite{stable-baselines} agent outperform DJIA in annual return and Sharpe ratio, as shown in Fig. \ref{fig:stock_trading} and Table \ref{table_stock}. The ElegantRL agent achieves an annual return of 22.425\% and a Sharpe ratio of $1.457$. The Stable-baselines3 agent achieves an annual return of 32.106\% and a Sharpe ratio of $1.621$. In the paper trading stage, the results are consistent with the backtesting results. Both the ElegantRL agent and the Stable-baselines3 agent outperform the baseline.

\begin{figure*}
\centering
\includegraphics[width=0.45\textwidth]{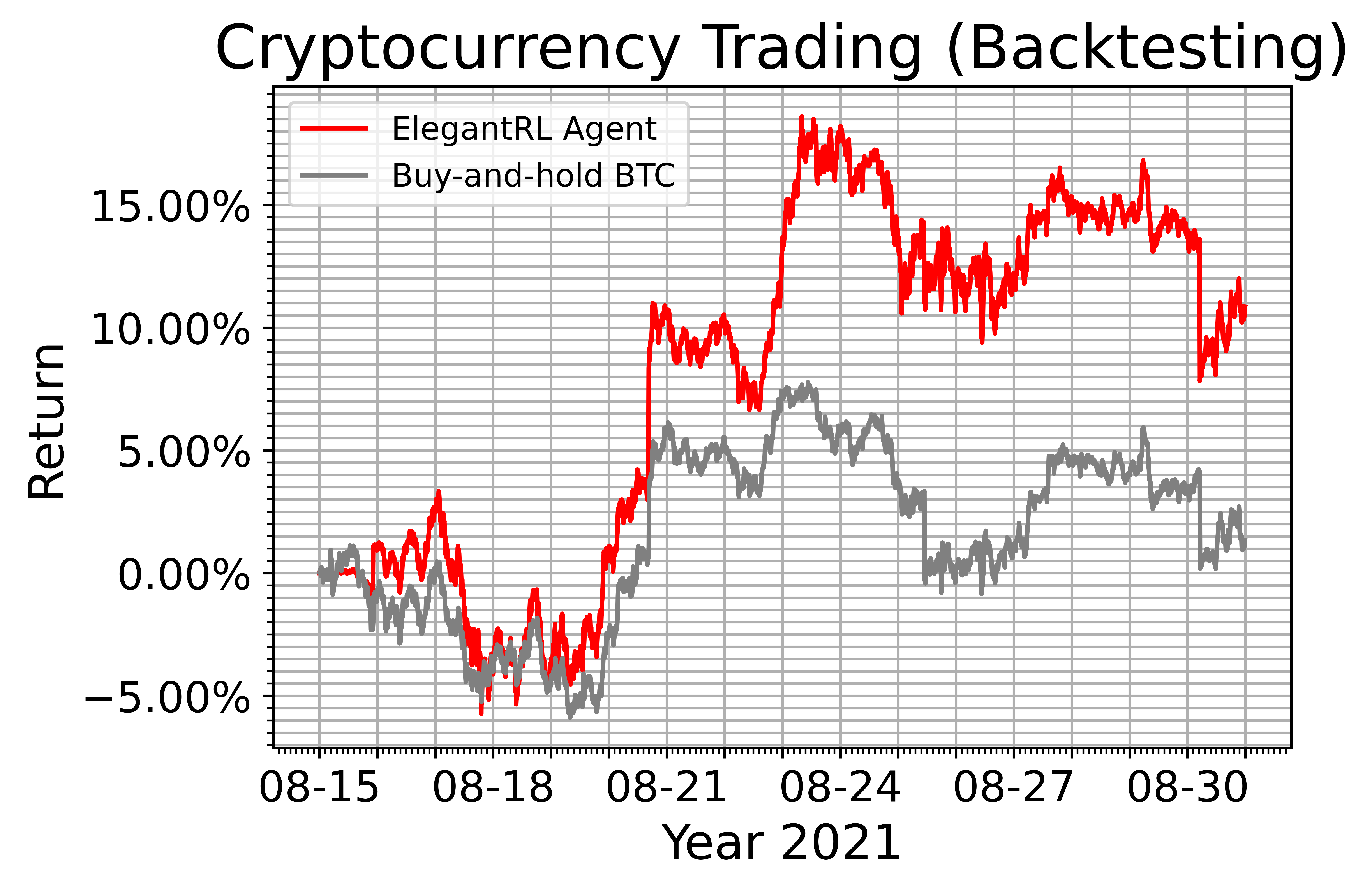}
\includegraphics[width=0.465\textwidth]{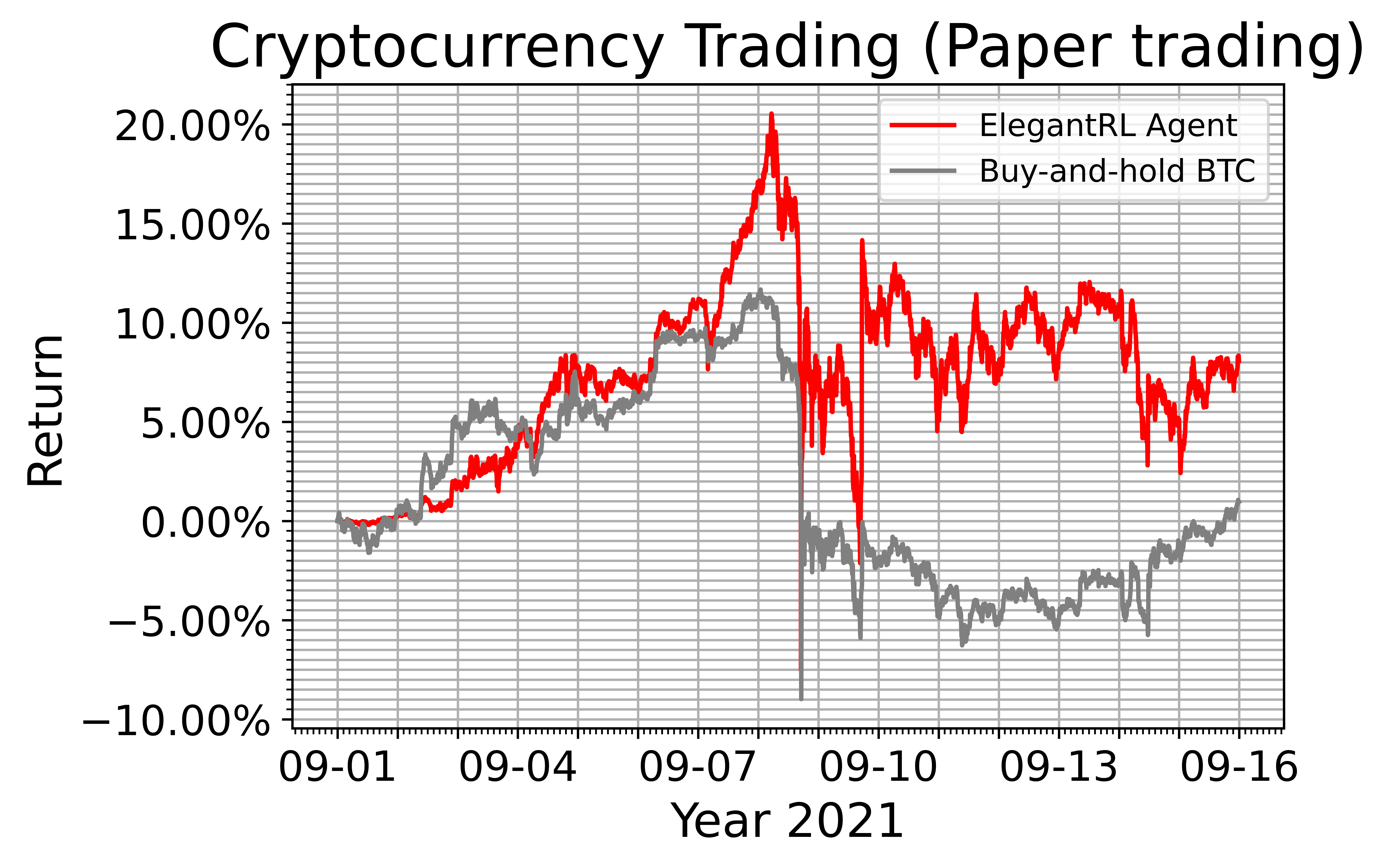}
\caption{Cumulative returns (5-minute) of cryptocurrency trading in backtesting and paper trading.}
\label{fig:crypto_trading}

\end{figure*}

\textbf{Cryptocurrency trading}:
In the backtesting stage, the ElegantRL agent outperforms the benchmark (BTC price) in most performance metrics, as shown in Fig. \ref{fig:crypto_trading} and Table \ref{table_crypto}. It achieves an annual return of 360.823\% and a Sharpe ratio of $2.992$. The ElegantRL agent also outperforms the benchmark (BTC price) in the paper trading stage, which is consistent with the backtesting results.


\begin{table*}[t]
\centering
\begin{tabular} {|l|c|c|} 
\hline
&\textbf{ElegantRL} \cite{elegantrl} & \textbf{BTC buy and hold}\\
\hline
\textbf{Cumul. return}  &\textcolor{red}{10.857\%} / \textcolor{blue}{4.844\%} & \textcolor{red}{1.332\%} / \textcolor{blue}{-1.255\%}\\
\hline
\textbf{Annual return}  &\textcolor{red}{360.823\%} / \textcolor{blue}{121.380\%} & \textcolor{red}{21.666\%} / \textcolor{blue}{5.492\%} \\
\hline
\textbf{Annual volatility} & \textcolor{red}{59.976\%} / \textcolor{blue}{65.857\%} & \textcolor{red}{47.410\%} / \textcolor{blue}{57.611\%} \\
\hline
\textbf{Sharpe ratio}  &     \textcolor{red}{2.992} / \textcolor{blue}{1.608} &   \textcolor{red}{0.657} / \textcolor{blue}{-0.113} \\
\hline
\textbf{Max drawdown} & \textcolor{red}{-6.396\%} / \textcolor{blue}{-10.474\%} & \textcolor{red}{-7.079\%} / \textcolor{blue}{-14.849\%}\\
\hline
\end{tabular}
\caption{Performance of backtesting  (\textcolor{red}{red}) and paper trading (\textcolor{blue}{blue}) for cryptocurrency trading.}
\label{table_crypto}
\vspace{-0.15in}
\end{table*}

\section{Conclusions}

In this paper, we followed the DataOps paradigm and developed a FinRL-Meta framework. FinRL-Meta provides open-source data engineering tools and hundreds of market environments with multiprocessing simulation.

For future work, we are building a multi-agent based market simulator that consists of over ten thousands of agents, namely, a FinRL-Metaverse. First, FinRL-Metaverse aims to build a universe of market environments, like the Xland environment \cite{deepmind2021open} and planet-scale climate forecast \cite{Ravuri2021} by DeepMind. To improve the performance for large-scale markets, we will employ GPU-based massive parallel simulation as Isaac Gym \cite{makoviychuk2021isaac}. Moreover, it will be interesting to explore the evolutionary perspective \cite{gupta2021embodied}\cite{scholl2021market}\cite{finrl_podracer_2021}\cite{gpu_podracer_nips_2021} to simulate the markets. We believe that FinRL-Metaverse will provide insights into complex market phenomena and offer guidance for financial regulations.
\section*{Acknowledgement}
This research used computational resources of the GPU cloud platform \cite{NVIDIA_SupPod2020} provided by the IDEA Research institute.

\medskip
\small
\bibliographystyle{plain}
\bibliography{ref}

\end{document}